\documentclass[preprint,12pt]{elsarticle}

\usepackage{graphicx}

\usepackage{amssymb}
\usepackage{amsmath}

\usepackage{lineno}

\journal{Nuclear Instruments and Methods B}

\begin{document}

\begin{frontmatter}



\title{Radiation of non-relativistic particle on a conducting sphere and a string of spheres}


\author[Kharkov]{N.F. Shul'ga}
\author[Belgorod]{V.V. Syshchenko\corref{sysh}}
\ead{syshch@yandex.ru}
\author[Belgorod]{E.A. Larikova}

\cortext[sysh]{Corresponding author. Tel.: +7 4722 301819; fax: +7 4722 301012}

\address[Kharkov]{A.I. Akhiezer Institute for Theoretical Physics, NSC ``KIPT'',\\ Akademicheskaya Street, 1, Kharkov 61108, Ukraine}
\address[Belgorod]{Belgorod State University, Pobedy Street, 85, Belgorod 308015, Russian Federation}

\begin{abstract}
The radiation resulting from the uniform motion of a charged particle by (or through) metal sphere is considered. The simple but rigorous description of the radiation process is developed for the case of non-relativistic particle and perfectly conducting sphere by the way of the method of images known from electrostatics. The spectral-angular and spectral densities of the diffraction and transition radiation on the single sphere are computed. The Smith-Purcell radiation caused by motion of the particle parallel to the periodic string of spheres is also considered.
\end{abstract}


\end{frontmatter}



\section{Introduction}
\label{intro}
The radiation emitted by a charged particle crossing the boundary between two media with different electrodynamic properties is known as transition radiation (TR), whereas for the particle traveling near the boundary of the spatially localized target without crossing the emitted radiation is called as diffraction radiation (DR). DR and TR of a charge on a perfectly conducting sphere (as well as on a periodic string of the spheres) are studied in the present paper.

Various aspects of DR and TR on spherical targets had been considered in numerous papers, see, e.g., \cite{Garcia}--\cite{Astap}. However, these papers either deal with some special cases or present the result in rather complicated form. Here we propose the simple and economic method for computation of the radiation characteristics (spectral-angular density as well as polarization, if needed) based on the following idea.

One of the ways to describe these types of radiation is the application of the boundary conditions to the Maxwell equations solutions for the field of the moving particle in two media. It becomes evident that the boundary conditions could be satisfied only after addition the solution of free Maxwell equations that corresponds to the radiation field, see, e.g. \cite{Ter.Mik}.

The conditions on the boundary between vacuum and ideal conductor could be satisfied in some cases via introduction of one or more fictitious charges along with the real charged particle; this approach to electrostatic problems is known as the method of images, see, e.g., \cite{Jackson}. The method of images had been used in the pioneering paper \cite{Ginz.Frank} where TR on a metal plane had been predicted. The method of images had been used also in \cite{Askaryan} for consideration of TR under passage of the particle through the center of the ideally conducting sphere in dipole approximation.

\section{DR on sphere}
\label{DR.section}
\begin{figure}
\begin{center}
\includegraphics[width=0.75\columnwidth]{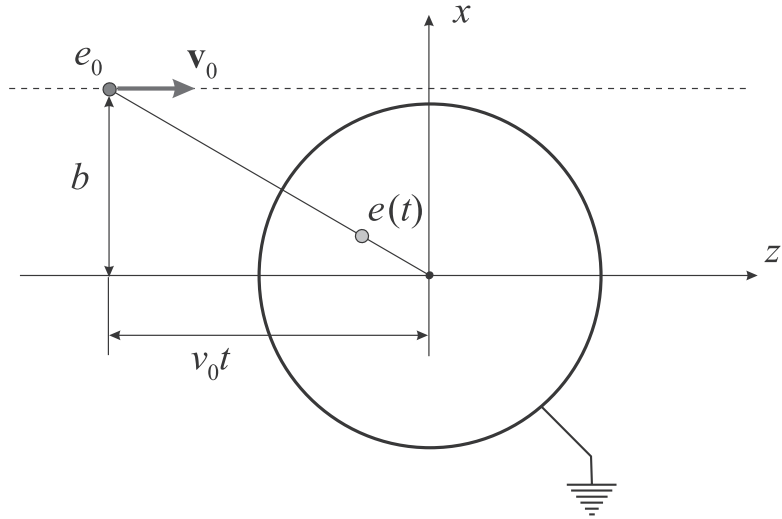}
\end{center}
\caption{Positions of the real charge $e_0$ and its ``image'' $e(t)$ near grounded conducting sphere of the radius $R$.}\label{Fig1}
\end{figure}

Consider the real charge $e_0$ passing near the grounded conducting sphere of the radius $R$ with the constant velocity $v_0 \ll c$, see Fig. \ref{Fig1}. Its image of the magnitude
\begin{equation}\label{fiktiv.1}
e(t) = -e_0 \frac{R}{\sqrt{b^2 + v_0^2 t^2}},
\end{equation}
has to be placed at the point with coordinates
\begin{equation}\label{fiktiv.2}
x(t) = \frac{R^2b}{b^2 + v_0^2 t^2}, \ z(t) = \frac{R^2 v_0 t}{b^2 + v_0^2 t^2}.
\end{equation}
So, while the incident particle moves uniformly, its ``image'' will move accelerated,
\begin{equation}\label{fiktiv.3}
v_x = \frac{dx}{dt} = - \frac{2R^2bv_0^2t}{(b^2 + v_0^2 t^2)^2},
 \ \ v_z = \frac{dz}{dt} = \frac{R^2 v_0 (b^2 - v_0^2 t^2)}{(b^2 + v_0^2 t^2)^2}.
\end{equation}
The radiation produced by non-uniform motion of the fictitious charge will be described by the well-known formula \cite{AhSh}
\begin{equation}\label{spectral.angular.0}
\frac{d\mathcal E}{d\omega d\Omega} = \frac{1}{4\pi^2c} \left|\mathbf k \times \mathbf I^{\vphantom{2}} \right|^2 ,
\end{equation}
where $\mathbf k$ is the wave vector of the radiated wave, $|\mathbf k| = \omega/c$, and
\begin{equation}\label{vector.I}
\mathbf I = \int_{-\infty}^\infty e(t)\,\mathbf v(t)\, e^{i(\omega t - \mathbf k\mathbf r(t))} \, dt
\end{equation}
(it could be easily seen that it is applicable to the case of time-varying charge $e(t)$ as well as to the case of the constant one). Note that the method of images for the isolated (in contrast to grounded) sphere requires introducing another fictitious charge of the magnitude $-e(t)$ resting at the center of the sphere. However, the last equation shows that such rest charge does not produce any radiation.

\begin{figure}
\begin{center}
\includegraphics[width=0.5\columnwidth]{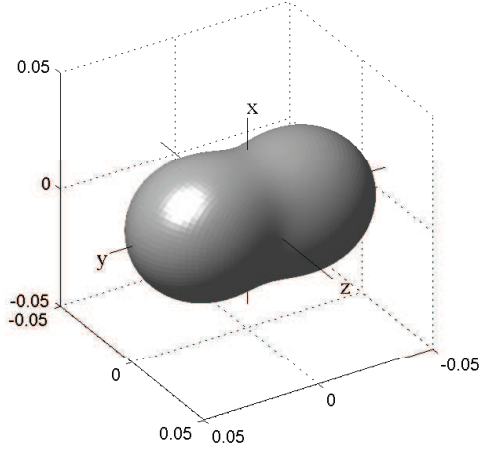}
\end{center}
\caption{The angular dependence of DR intensity as direction diagram for the passage of the real charge under b = R (sliding incidence on the sphere, when DR intensity is maximal for the whole range of wavelengths) and $R\omega/v_0 = 2.34$ (this choice is due to the maximum of DR spectrum (see below) falls on $b\omega/v_0 \approx 2.34$ and $b = R$ in the given case).}\label{Fig2}
\end{figure}

Substitution of (\ref{fiktiv.1}), (\ref{fiktiv.2}), (\ref{fiktiv.3}) into (\ref{vector.I}) gives the following integrals:
\begin{equation}\label{I.x.1}
I_x = 2e_0 b R^3 v_0^2 \times
\end{equation}
$$
\times \int_{-\infty}^\infty
\exp\left\{i\left[ \omega t - \frac{k_x R^2b}{b^2 + v_0^2 t^2} - \frac{k_z R^2 v_0t}{b^2 + v_0^2 t^2}\right] \right\}
\frac{t\,dt}{(b^2 + v_0^2 t^2)^{5/2}} ,
$$
\begin{equation}\label{I.z.1}
I_z = -e_0 R^3 v_0 \times
\end{equation}
$$
\times \int_{-\infty}^\infty
\exp\left\{i\left[ \omega t - \frac{k_x R^2b}{b^2 + v_0^2 t^2} - \frac{k_z R^2 v_0t}{b^2 + v_0^2 t^2}\right] \right\}
\frac{(b^2 - v_0^2 t^2)\,dt}{(b^2 + v_0^2 t^2)^{5/2}} .
$$
The integration in (\ref{I.x.1}), (\ref{I.z.1}) can be easily performed numerically, that leads to the spectral-angular density of diffraction radiation in the form
\begin{equation}\label{spectral.angular.1}
  \frac{d\mathcal E}{d\omega d\Omega} = \frac{e_0^2}{4\pi^2c} \, \Phi_{DR}(\theta , \varphi , \omega) ,
\end{equation}
where the typical shape of the angular distribution $\Phi_{DR}(\theta , \varphi , \omega)$ is presented in Fig. \ref{Fig2}; for different values of parameters it does not vary too much.

Approximate analytical result can be obtained after neglecting the second term in the exponents in (\ref{I.x.1}), (\ref{I.z.1}) (that is valid for low radiation frequencies, $\omega \ll cb/R^2$) as well as the third term (that is always valid for non-relativistic case):
$$
\frac{d\mathcal E}{d\omega d\Omega} = \frac{e_0^2}{4\pi^2c} \, \frac{16 R^6 \omega^6}{9c^2 v_0^4}
\left\{ \left( 1 - \sin^2\theta \cos^2\varphi \right) K_1^2 \left(\frac{\omega}{v_0} b \right) + \right.
$$
\begin{equation}\label{spectral.angular.2}
\left. +
\sin^2\theta \left(\frac{v_0}{2\omega\,b}\right)^2 \left[ K_1 \left(\frac{\omega}{v_0} b \right) +
2 \frac{\omega}{v_0}\,b\, K_0 \left(\frac{\omega}{v_0} b \right) \right]^2 \right\},
\end{equation}
where $K_0$ and $K_1$ are modified Bessel functions of the third kind.

\begin{figure}
\begin{center}
\includegraphics[width=0.75\columnwidth]{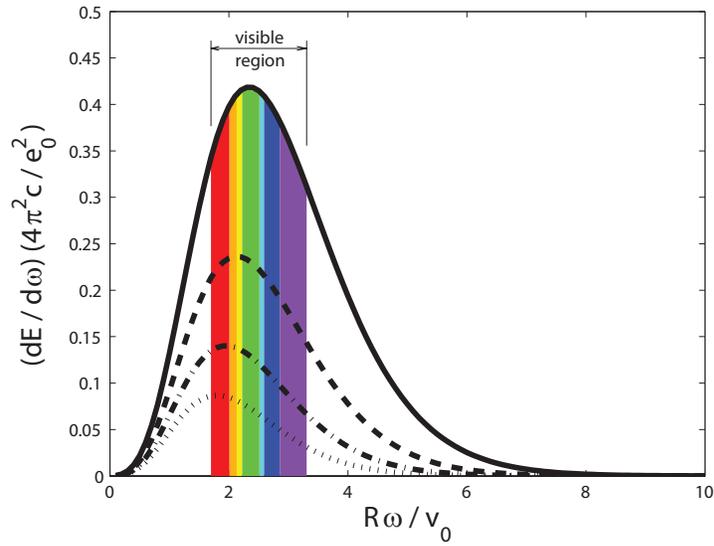}
\end{center}
\caption{DR spectrum for $R = 20$ nm, $v_0 = 0.1c$ and $b = R$ (solid line), $b = 1.1R$ (dashed line), $b = 1.2R$ (dash-dotted line), $b = 1.3R$ (dotted line).}\label{Fig3}
\end{figure}

Integration of (\ref{spectral.angular.1}) over radiation angles leads to the radiation spectrum presented in Fig. \ref{Fig3}. For illustrative purposes, we choose the parameters $v_0 = 0.1c$, $R = 20$ nm, $b = R+0$, for which the DR intensity maximum will lie in the visible spectrum. Note, however, that in this frequency domain the properties of the sphere material can be far from that of a perfect conductor. Particularly, plasma oscillations can be important. On the other side, the perfect conductor approximation is valid for the frequencies less than the inverse relaxation time $\tau^{-1}$ for the electrons in the metal. For instance, $\tau^{-1} = 5\cdot 10^{-13}$ s$^{-1}$ for copper \cite{Kittel}, so the results obtained surely could be applied up to THz and far infrared range. The applicability of the results to higher frequencies needs further investigations.

\section{DR on the string of spheres}
\label{DR.string.section}
Now consider the motion of a charge $e_0$ along the periodic string of $N\gg 1$ spheres. The mutual influence of the fictitious charges induced in the neighboring spheres can be neglected in the case of small impact parameters, $b\to R$ (when the radiation intensity is high), for the string period large enough, $a\gtrsim 5R$. Then the interference of the radiation produced on the subsequent spheres leads to the simple formula for the spectral-angular density of DR:
\begin{equation}\label{spectral.angular.string.1}
  \frac{d\mathcal E}{d\omega d\Omega} = \frac{e_0^2}{4\pi^2c} \, \Phi_{DR}(\theta , \varphi , \omega) \times
\end{equation}
$$
\times 2\pi N \frac{v_0}{\omega a} \sum_{m=1}^\infty \delta \left(1 - \frac{v_0}{c} \cos\theta - m \frac{2\pi v_0}{\omega a} \right) ,
$$
where the delta-function means well-known Smith--Purcell condition \cite{Smith.Purcell}.

\begin{figure}
\includegraphics[width=0.49\columnwidth]{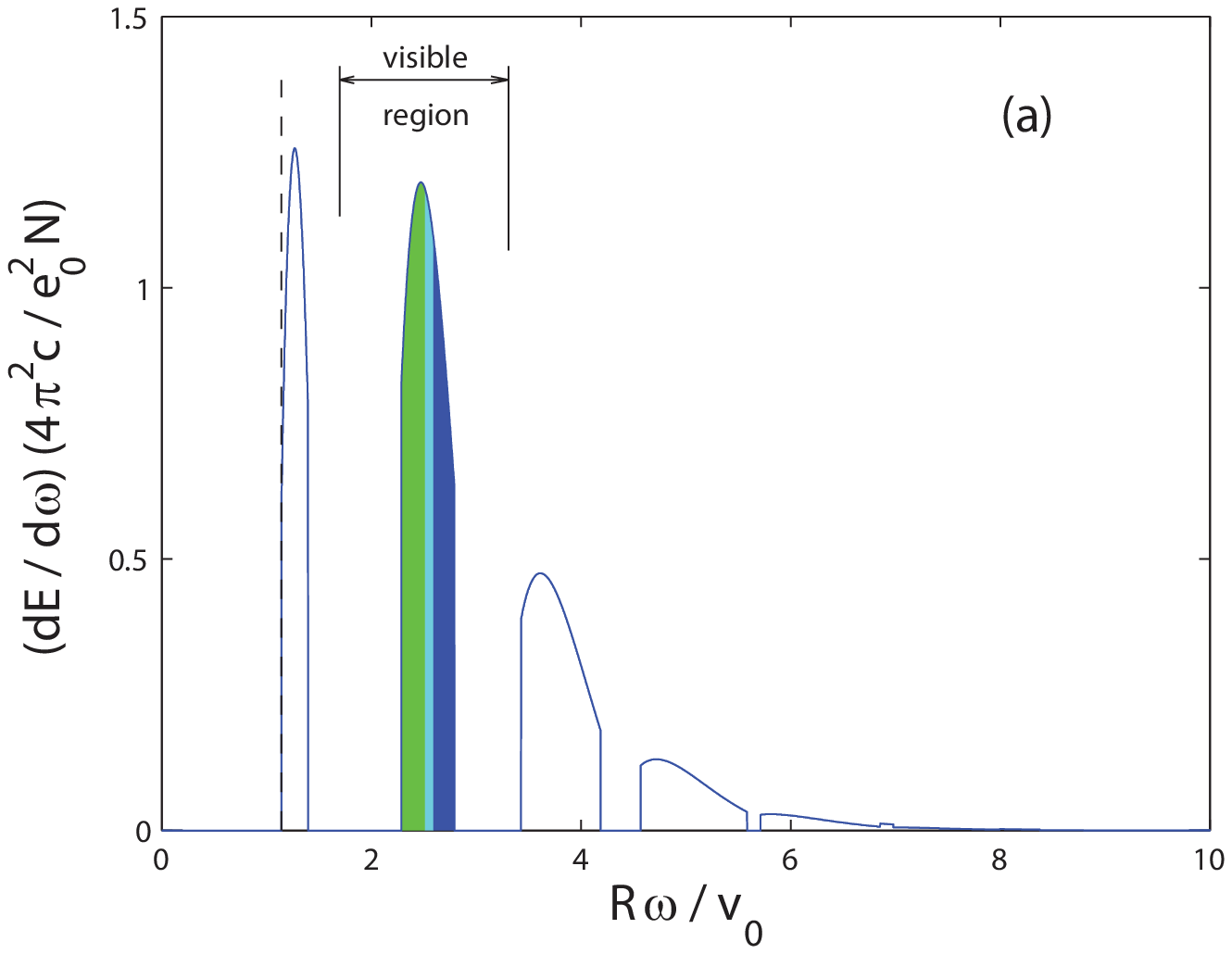} \ \includegraphics[width=0.49\columnwidth]{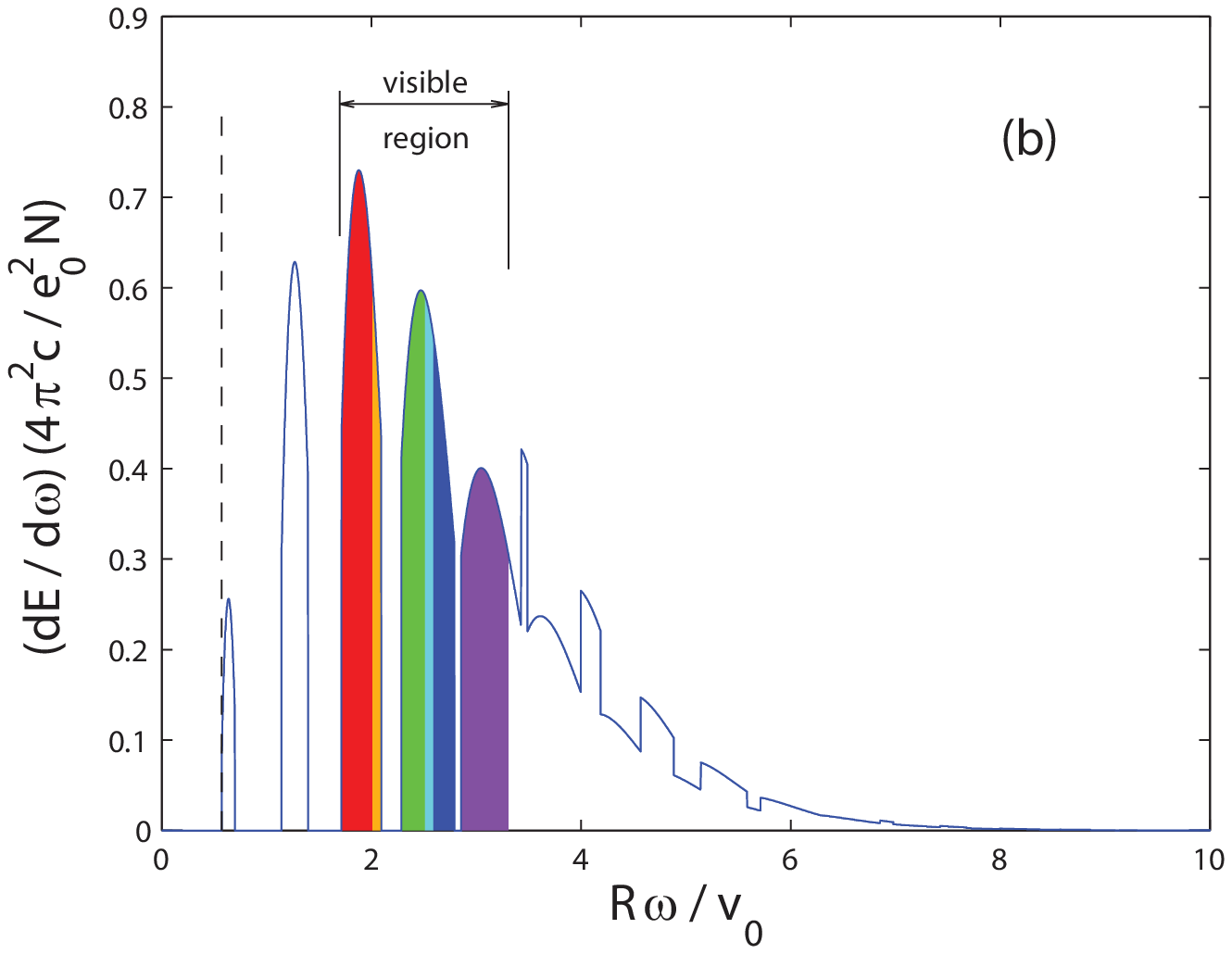} \\
\includegraphics[width=0.49\columnwidth]{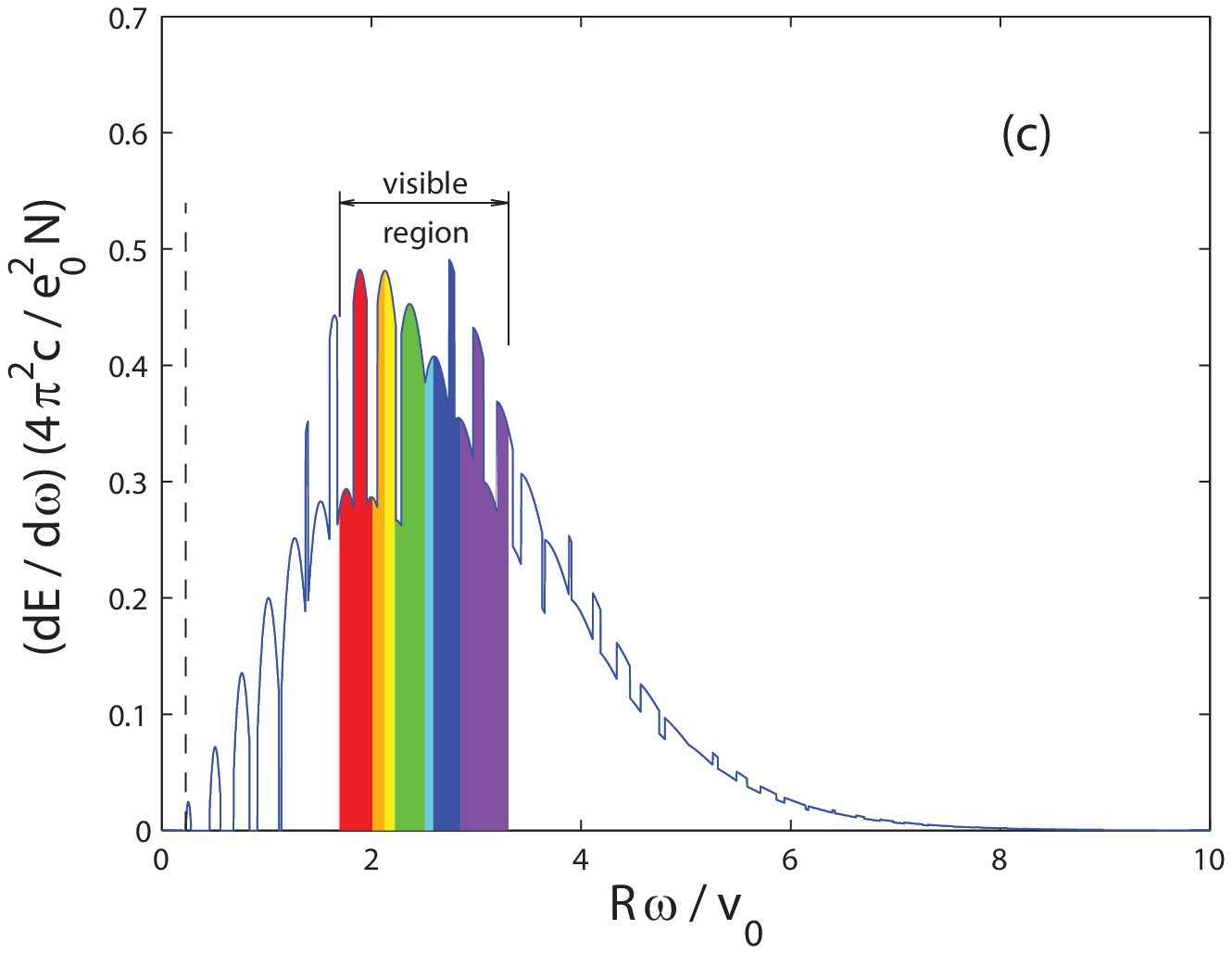} \ \includegraphics[width=0.49\columnwidth]{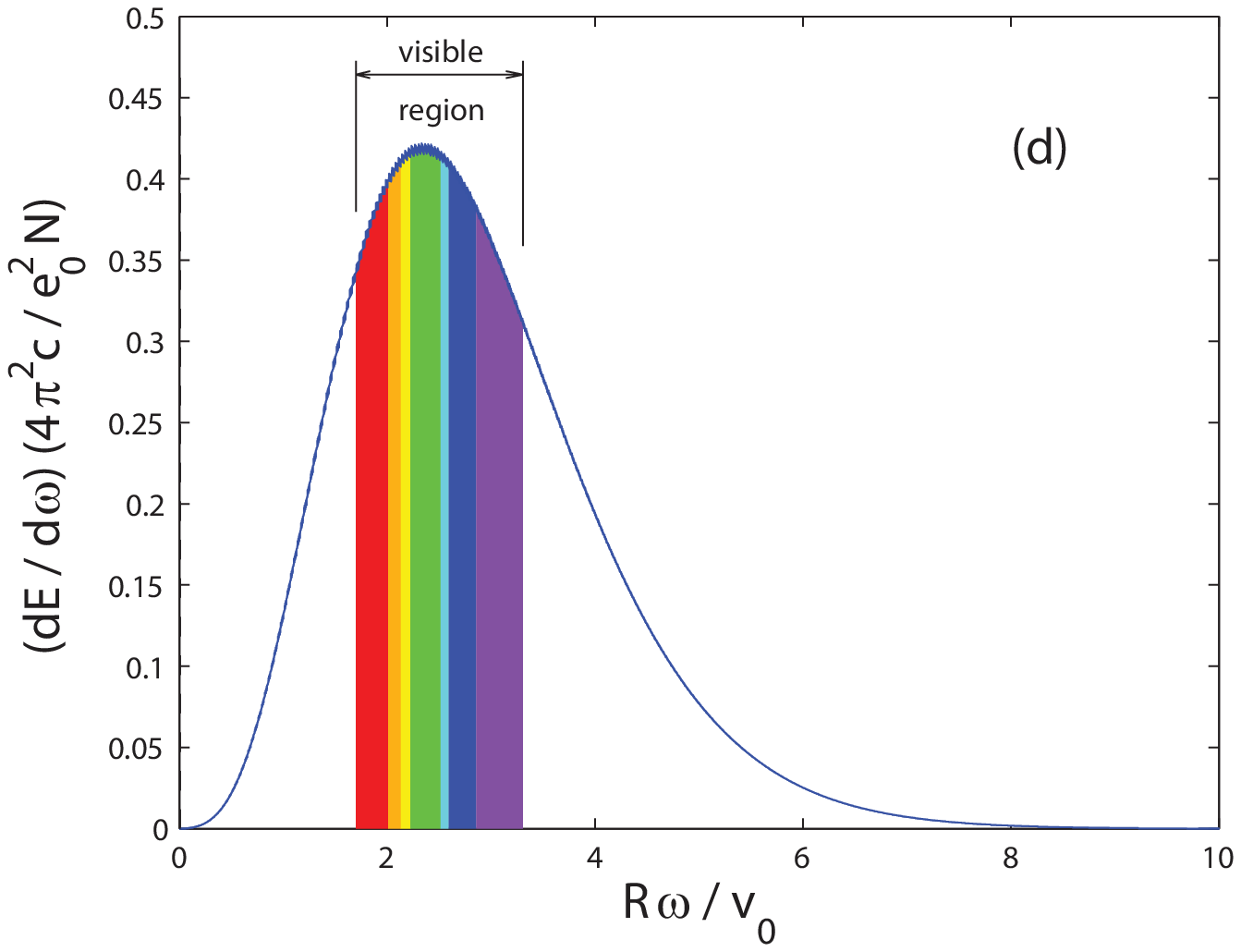} \\
\caption{DR spectrum on the string of spheres under $b = R$, $R = 20$ nm, $v_0 = 0.1c$, $a = 100$, 200, 500, 20000 nm. Vertical dashed line marks the long wavelength edge of the spectrum.}\label{Fig4}
\end{figure}

The spectrum for the string period $a$ small enough consists of separated bands, see Fig. \ref{Fig4}, a. The bands overlap each other under increase of the string period $a$ gradually forming the spectrum of DR on a single sphere (multiplied by the total number of the spheres $N$), see Fig. \ref{Fig4}, b, c, d.

\section{TR on the sphere}
\label{TR.section}
\begin{figure}
\includegraphics[width=0.49\columnwidth]{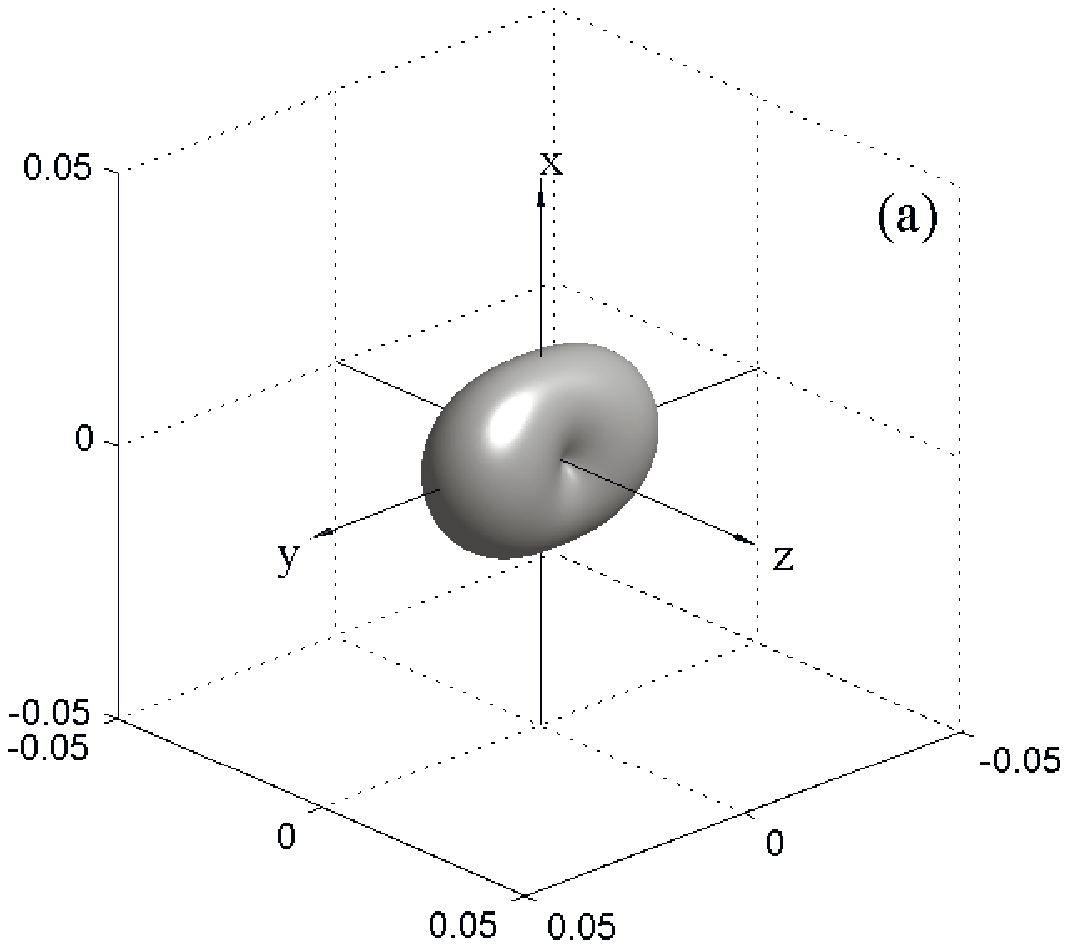} \ \includegraphics[width=0.49\columnwidth]{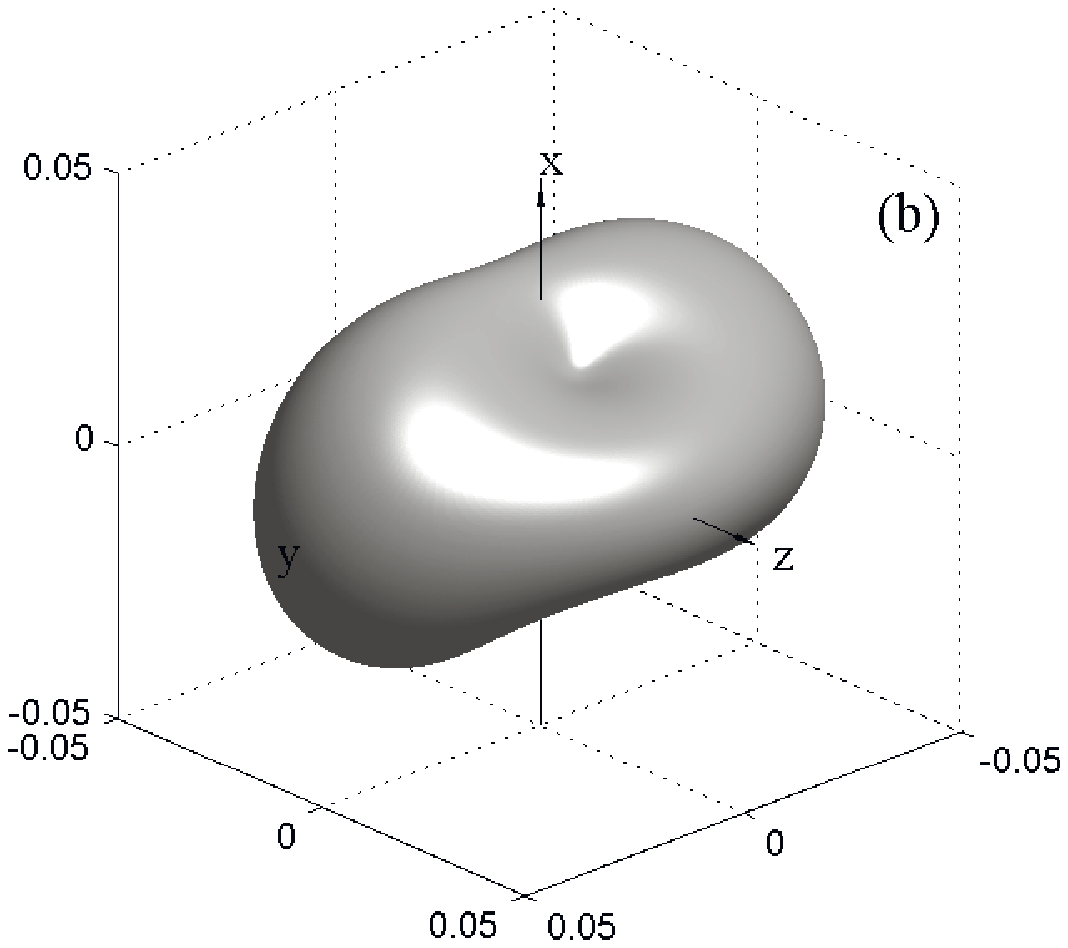} \\
\includegraphics[width=0.49\columnwidth]{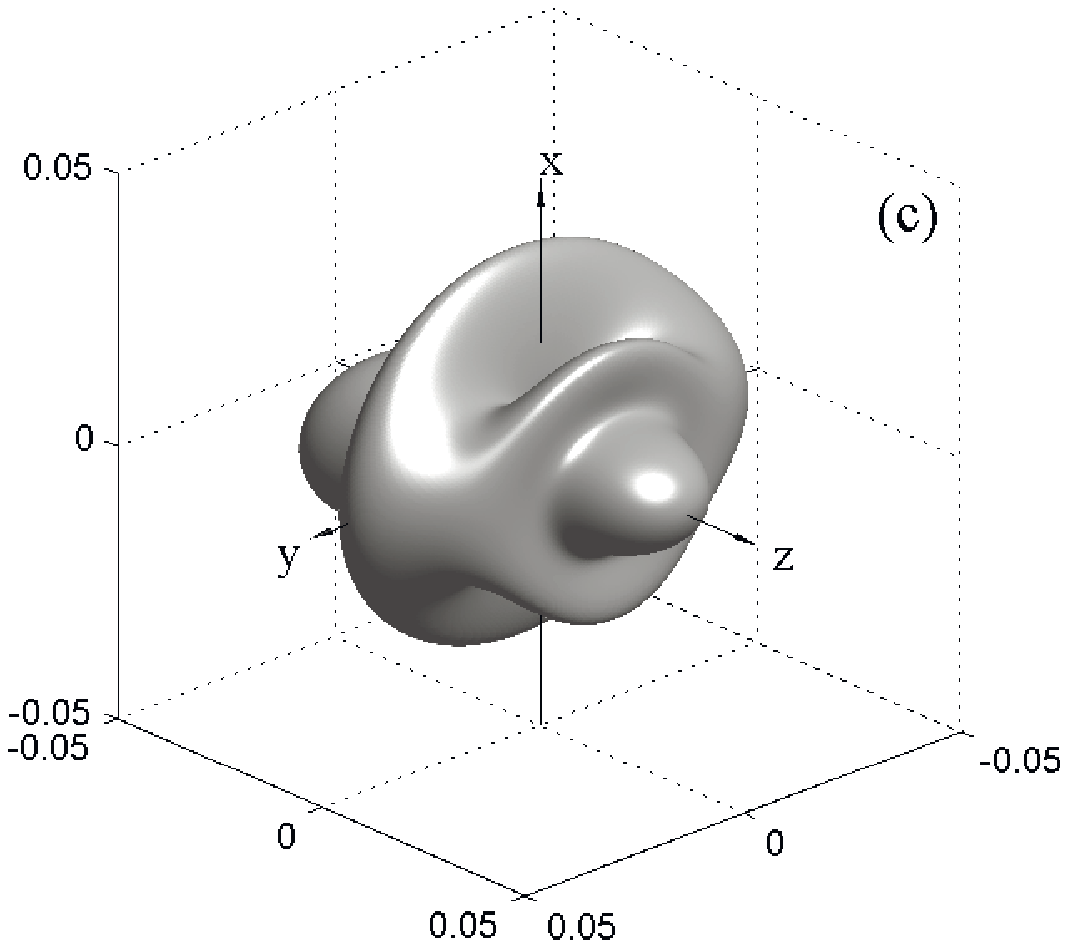} \ \includegraphics[width=0.49\columnwidth]{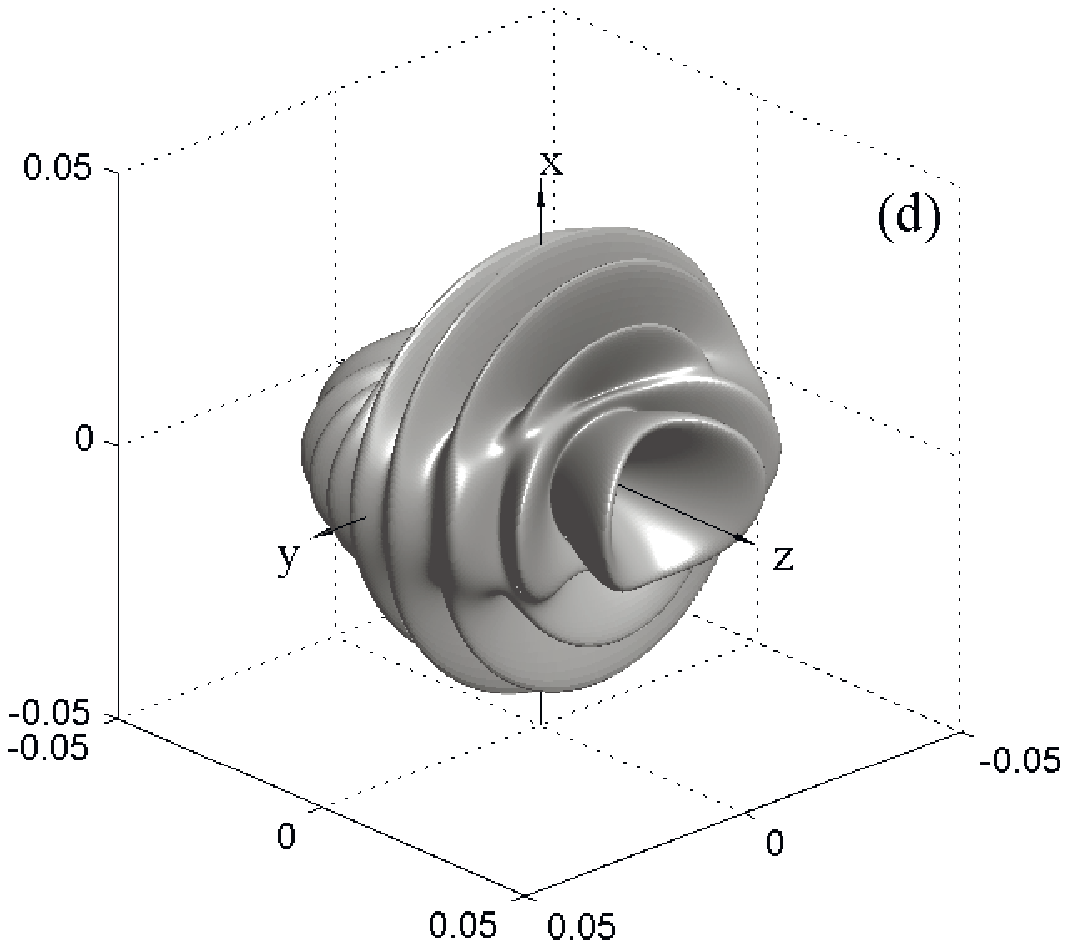} \\
\caption{Direction diagrams of the radiation emitted under incidence of the real charge under $b=R/\sqrt{2}$ for $v_0 = 0.1c$, $R\omega/v_0 = 1$, 10, 50, 200.}\label{Fig5}
\end{figure}

TR arises under $b < R$, when the incident charge $e_0$ crosses the sphere. In this case both real and fictitious charges vanish while crossing the sphere (and then appear again) that complicates the formulae describing the radiation. For the convenience of the further interpretation, let us separate the contributions into the vector $\mathbf I$ (\ref{vector.I}) from the real particle and its image on the time interval $-\infty < t \leq -t_0$, where $t_0 = \sqrt{R^2-b^2}/v_0$, that is on the incoming part of the particle's trajectory,
\begin{equation}\label{TR.I.z.real.in}
I_z^{(real.in)} = -ie_0v_0 \frac{\exp[-ik_xb-i(\omega - k_zv_0)t_0]}{\omega - k_zv_0} ,
\end{equation}
\begin{equation}\label{TR.I.x.image.in}
I_x^{(image.in)} = 2e_0 b R^3 v_0^2 \times
\end{equation}
$$
\times \int_{-\infty}^{-t_0}
\exp\left\{i\left[ \omega t - \frac{k_x R^2b}{b^2 + v_0^2 t^2} - \frac{k_z R^2 v_0t}{b^2 + v_0^2 t^2}\right] \right\}
\frac{t\,dt}{(b^2 + v_0^2 t^2)^{5/2}} ,
$$
\begin{equation}\label{TR.I.z.image.in}
I_z^{(image.in)} = -e_0 R^3 v_0 \times
\end{equation}
$$
\times \int_{-\infty}^{-t_0}
\exp\left\{i\left[ \omega t - \frac{k_x R^2b}{b^2 + v_0^2 t^2} - \frac{k_z R^2 v_0t}{b^2 + v_0^2 t^2}\right] \right\}
\frac{(b^2 - v_0^2 t^2)\,dt}{(b^2 + v_0^2 t^2)^{5/2}} ,
$$
and on the interval $t_0 \leq t < \infty$, that is on the outgoing part of the particle's trajectory,
\begin{equation}\label{TR.I.z.real.out}
I_z^{(real.out)} = ie_0v_0 \frac{\exp[-ik_xb+i(\omega - k_zv_0)t_0]}{\omega - k_zv_0} ,
\end{equation}
\begin{equation}\label{TR.I.x.image.out}
I_x^{(image.out)} = 2e_0 b R^3 v_0^2 \times
\end{equation}
$$
\times \int_{t_0}^\infty
\exp\left\{i\left[ \omega t - \frac{k_x R^2b}{b^2 + v_0^2 t^2} - \frac{k_z R^2 v_0t}{b^2 + v_0^2 t^2}\right] \right\}
\frac{t\,dt}{(b^2 + v_0^2 t^2)^{5/2}} ,
$$
\begin{equation}\label{TR.I.z.image.out}
I_z^{(image.out)} = -e_0 R^3 v_0 \times
\end{equation}
$$
\times \int_{t_0}^\infty
\exp\left\{i\left[ \omega t - \frac{k_x R^2b}{b^2 + v_0^2 t^2} - \frac{k_z R^2 v_0t}{b^2 + v_0^2 t^2}\right] \right\}
\frac{(b^2 - v_0^2 t^2)\,dt}{(b^2 + v_0^2 t^2)^{5/2}} .
$$
Numerical integration leads to the spectral-angular density of transition radiation in the form
\begin{equation}\label{spectral.angular.10}
  \frac{d\mathcal E}{d\omega d\Omega} = \frac{e_0^2}{4\pi^2c} \, \Phi_{TR}(\theta , \varphi , \omega) ,
\end{equation}
where the function $\Phi_{TR}(\theta , \varphi , \omega)$ is presented for some particular cases as the directional diagram in Fig. \ref{Fig5}. We see very sophisticate shapes in contrast to DR case. This is due to interference of the radiation emitted by the real charge and its image while crossing two boundaries between vacuum and conductor.

\begin{figure}
\includegraphics[width=0.49\columnwidth]{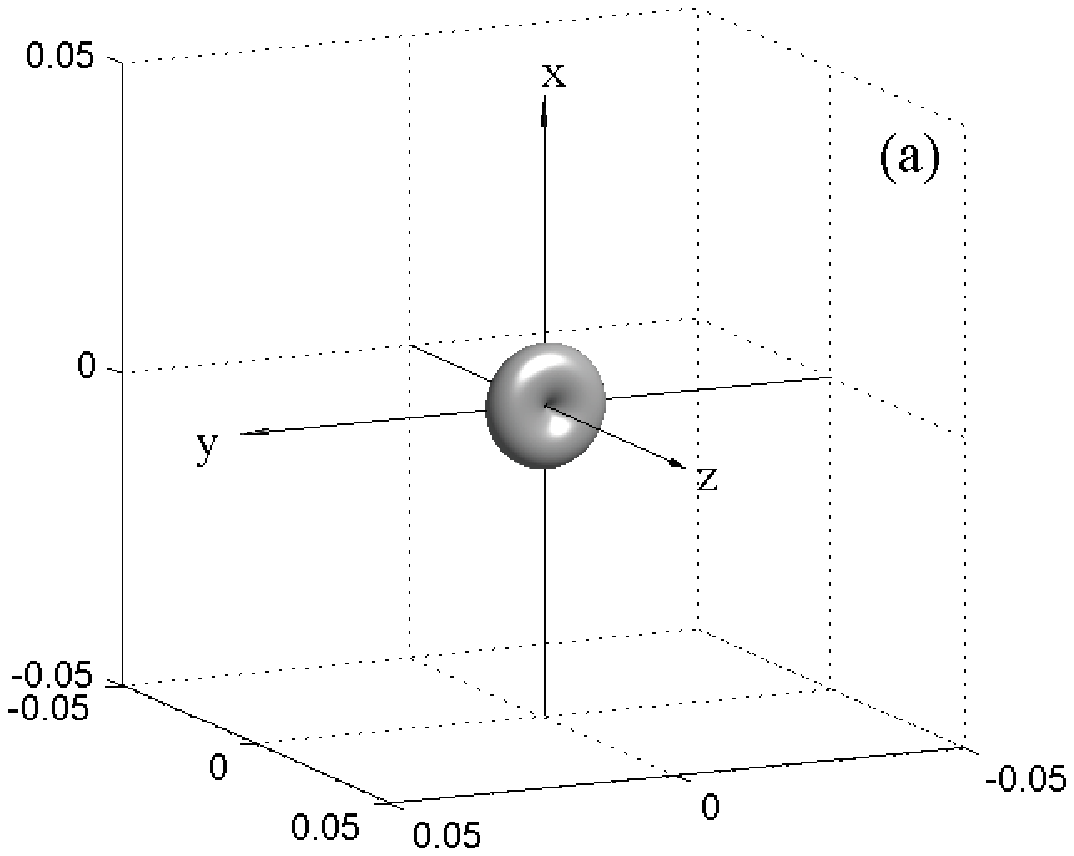} \ \includegraphics[width=0.49\columnwidth]{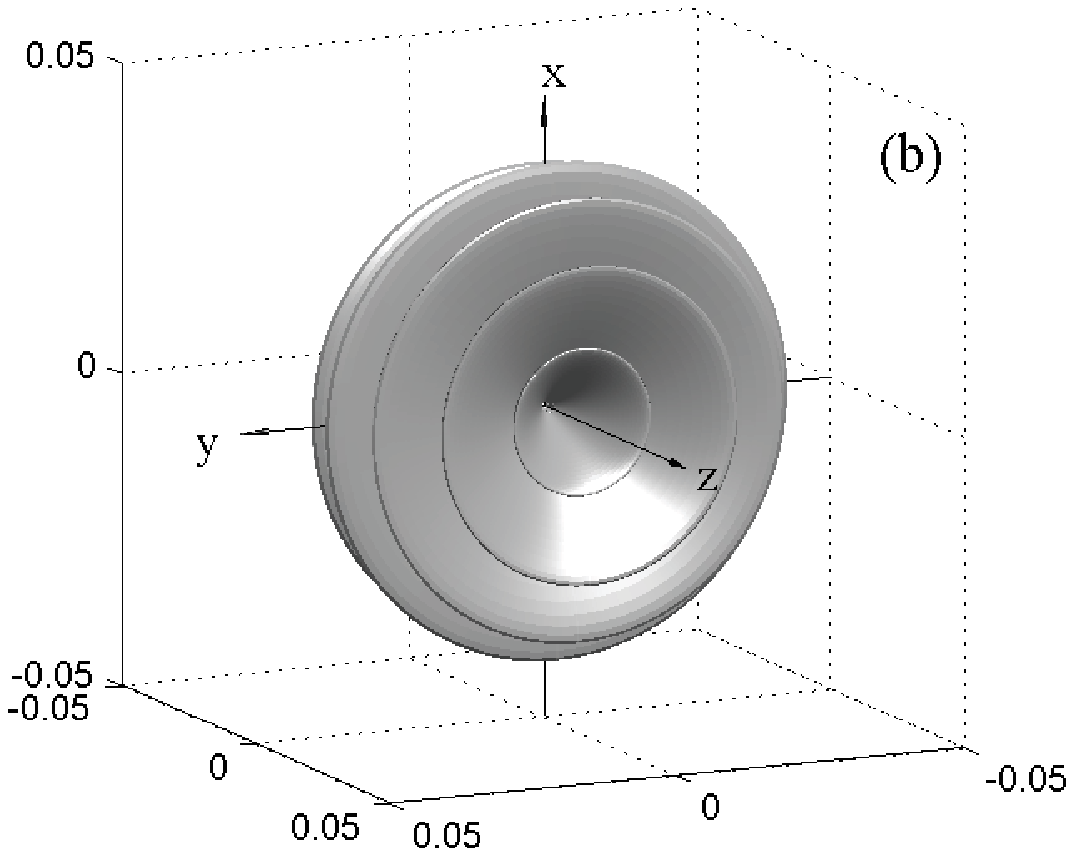} \\
\includegraphics[width=0.49\columnwidth]{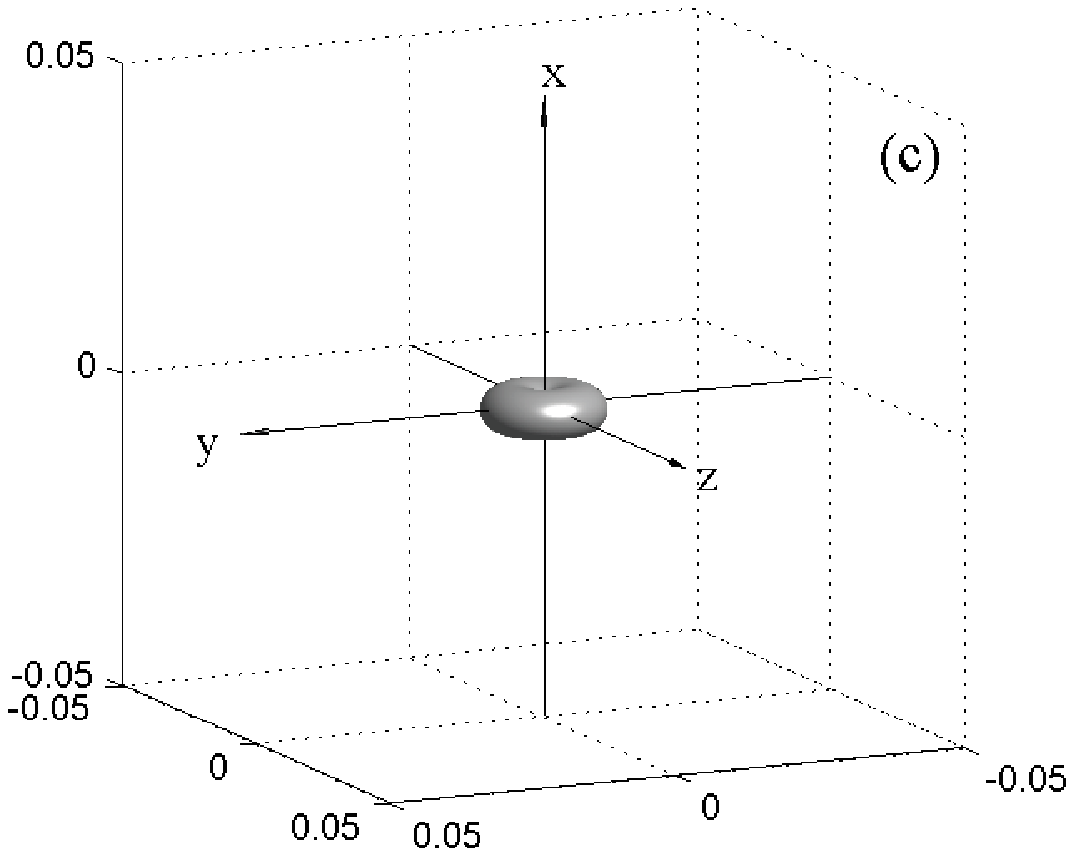} \ \includegraphics[width=0.49\columnwidth]{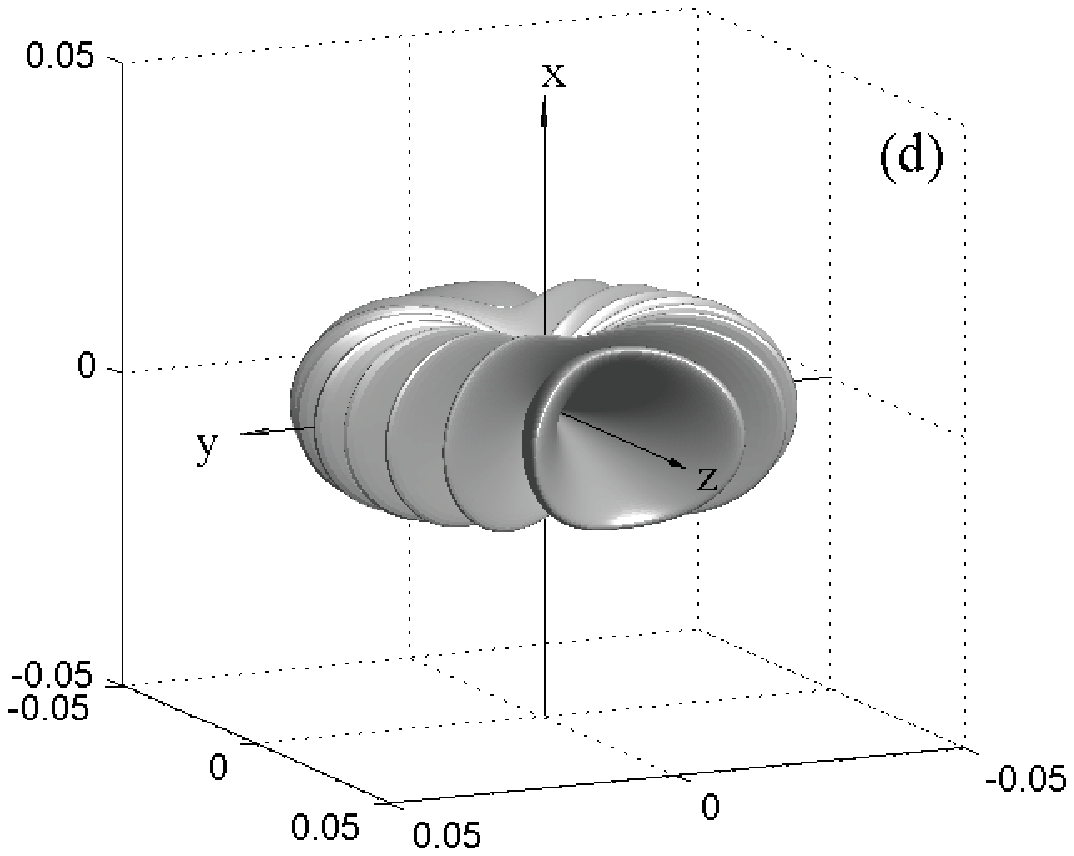} \\
\caption{Direction diagrams of the partial contributions to the radiation emitted under incidence of the real charge under $b=R/\sqrt{2}$ for $R\omega/v_0 = 200$, $v_0 = 0.1c$.}\label{Fig6}
\end{figure}

Let us trace out the origin of this interference picture in the case of high radiation frequency $R\omega/v_0\gg 1$, when the transverse size of the particle's Coulomb field Fourier component is small and the curvature of the conductor surface is negligible. The elementary contributions from both real and fictitious charges into the radiation under each crossing of the metal surface have similar shape with axis of symmetry along their velocity. So, for the real particle the axis of symmetry is directed along the $z$ axis, see Fig. \ref{Fig6}, a, where the directional diagram computed with the vector $\mathbf I^{(real.in)}$ (\ref{TR.I.z.real.in}) or $\mathbf I^{(real.out)}$ (\ref{TR.I.z.real.out}) instead of complete vector $\mathbf I$ is presented. The interference of two such contributions from the enter point of the charge into the sphere and from the exit one leads to the directional diagram in the Fig. \ref{Fig6}, b.

On the other hand, the axis of symmetry of the elementary contribution from the fictitious charge for $b=R/\sqrt{2}$ (that is for the incidence under 45 degrees) will be directed along the $x$ axis (Fig. \ref{Fig6}, c, which is computed for the vector $\mathbf I^{(image.in)}$ (\ref{TR.I.x.image.in}), (\ref{TR.I.z.image.in}) or $\mathbf I^{(image.out)}$ (\ref{TR.I.x.image.out}), (\ref{TR.I.z.image.out})). The interference of two such contributions from two points of crossing the sphere leads to the picture in the Fig. \ref{Fig6}, d.

Finally, the interference of the contributions from the real charge (Fig. \ref{Fig6}, b) and its image (Fig. \ref{Fig6}, d) leads to the final picture of the radiation emission (Fig. \ref{Fig5}, d).

\begin{figure}
\begin{center}
\includegraphics[width=0.75\columnwidth]{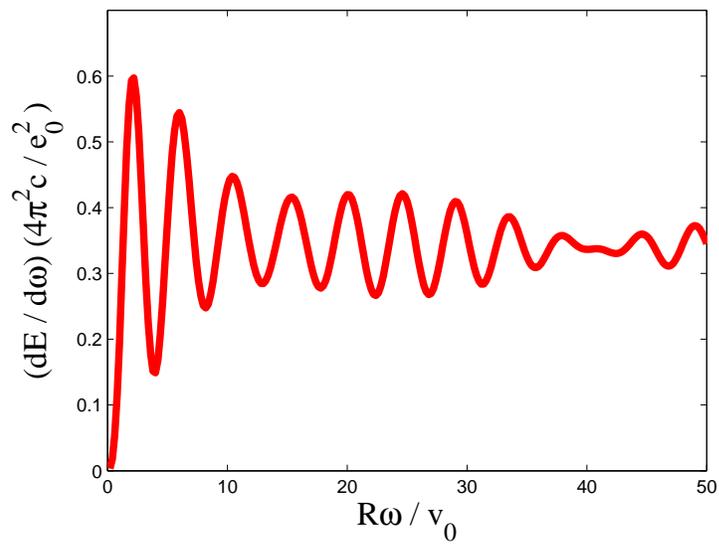}
\end{center}
\caption{TR spectrum (\ref{spectral.1}) for $b=R/\sqrt{2}$, $v_0 = 0.1c$.}\label{TR.spectr}
\end{figure}

TR spectral density
\begin{equation}\label{spectral.1}
  \frac{d\mathcal E}{d\omega} = \frac{e_0^2}{4\pi^2c} \, \int \Phi_{TR}(\theta , \varphi , \omega) \,  d\Omega
\end{equation}
for $b=R/\sqrt{2}$ is presented in Fig. \ref{TR.spectr}.

\section{Conclusion}
\label{conclusion}
The radiation resulting from the interaction of non-relativistic particle with perfectly conducting sphere is considered.

The method of images allows the precise description of the radiation in this case.
The integration of the resulting formulae can be easily performed numerically that permits to compute the spectral-angular density of the diffraction and transition radiation for an arbitrary impact parameter. The angular distribution of the transition radiation shows a wide diversity of shapes for different values of the radiation frequency and other parameters. This effect is the result of the interference of the waves emitted by the particle and its image at the moment when the boundary between the conductor and vacuum is being crossed by the particle.

The results are valid only for the non-relativistic particles due to the nature of the method of images: the sphere on which the potential of two charges (the real one and its image) is equal to zero exists only for spherically symmetric Coulomb potentials, not for the relativistically compressed ones. However, when the characteristic transverse size of the incident particle's Coulomb field $v_0/\omega$ becomes much less than the sphere radius $R$ and the impact parameter is not close to $R$, one can neglect the sphere's curvature. In this case our results for TR could be applied also for relativistic particles.

Our results are valid at least up to THz and far infrared domain. Their applicability for higher frequencies, where the sphere's material properties are far from ones of the perfect conductor, needs further investigations.

Note that the trajectory of the particle interacting with a sphere or even a string of spheres remains rectilinear with a high degree of accuracy providing that the particle is heavy, like a proton or a nucleus. One can minimize the perturbation of the particle's trajectory via using the spheres made of thin foil or drilling a fine channel through the spheres for the particle's flight.

\section{Acknowledgements}
\label{acknow}
The work was supported in part by the grant of Russian Science Foundation (project 15-12-10019).


\end{document}